\documentclass{INTERSPEECH2023}

\usepackage{graphicx} 
\graphicspath{ {./Images/} }
\usepackage{comment}
\usepackage{xcolor,colortbl}

\interspeechcameraready

\title{Why We Should Report the Details in Subjective Evaluation \\of TTS More Rigorously}
\name{Cheng-Han Chiang, Wei-Ping Huang, Hung-yi Lee}
\address{
  National Taiwan University, Taiwan}
\email{dcml0714@gmail.com, r11942102@ntu.edu.tw, hungyilee@ntu.edu.tw}

\begin{document}

\maketitle
 
\begin{abstract}
This paper emphasizes the importance of reporting experiment details in subjective evaluations and demonstrates how such details can significantly impact evaluation results in the field of speech synthesis. 
Through an analysis of 80 papers presented at INTERSPEECH 2022, we find a lack of thorough reporting on critical details such as evaluator recruitment and filtering, instructions and payments, and the geographic and linguistic backgrounds of evaluators. 
To illustrate the effect of these details on evaluation outcomes, we conducted mean opinion score (MOS) tests on three well-known TTS systems under different evaluation settings and we obtain at least three distinct rankings of TTS models. 
We urge the community to report experiment details in subjective evaluations to improve the reliability and interpretability of experimental results.

\end{abstract}
\noindent\textbf{Index Terms}: mean opinion score, naturalness, listening test, crowdsourcing, Amazon Mechanical Turk

\section{Introduction}
\label{section: Introduction}
Speech synthesis is the fundamental building block to several speech processing tasks, such as text-to-speech (TTS), voice conversion~\cite{wang22u_interspeech}, and speech-to-speech translation~\cite{liu22u_interspeech}. 
Due to the absence of ground truth and automatic evaluation metrics, subjective evaluation~\cite{rec1996p} is the predominant method used to assess the quality of synthesized speech.
In the subjective evaluation, researchers recruit listeners and present the listeners with some speech signals, and the listeners are asked to rate the given speech signal based on the task instructions given to the human evaluators.
Using online crowdsourcing platforms has been more and more common these days~\cite{rec2018p}.

Despite subjective evaluation being a critical evaluation metric for speech synthesis systems, we discover that prior works often omit details pertaining to subjective evaluation.
Through an analysis of over 80 papers presented at INTERSPEECH 2022 on speech synthesis, we find that none of the papers provide comprehensive details to enable the replication of subjective evaluation under the same experimental setting.
These missing details include the recruitment and selection of evaluators, their instructions and compensation, their qualifications, location, and linguistic background.

To show that these missing details in subjective evaluation can significantly influence the experiment result, we conduct mean opinion score (MOS) tests to assess the quality of three different TTS models: Tacotron2~\cite{https://doi.org/10.48550/arxiv.1712.05884}, FastSpeech2~\cite{https://doi.org/10.48550/arxiv.2006.04558}, and VITS~\cite{https://doi.org/10.48550/arxiv.2106.06103}.
We perform over ten sets of MOS tests on the quality of audio samples generated by the TTS models and ground truth human recordings, with the same audio samples used across all MOS tests. 
The MOS tests differ in some experiment details that are omitted in prior works.
Since all MOS tests we conduct share the same audio samples, we expect only one "ground truth ranking" on the quality of audio samples generated by different TTS models, but our MOS tests yield at least three rankings on the three TTS models.
Our results highlight the criticality of details in subjective evaluations for reliable experiment results.

\section{Survey of Prior Works}
\label{section: A Survey of INTERSPEECH 2022 Papers Involving Speech Synthesis and Subjective Evaluations}
We begin by conducting a survey of previous works to comprehend the current state of how the details in subjective evaluation experiments are reported.
Specifically, we survey \textbf{all} the papers in INTERSPEECH 2022 that belong to the speech synthesis track or have the term "speech synthesis" in the paper's title and conduct subjective evaluation.
We exclude 8 papers that do not use MOS evaluation, resulting in a total of 80 papers.
 For each of these papers, we evaluate whether they report the following \textbf{factors} or not:

\textbf{Recruitment platform:}
Out of the 80 papers examined, 62 do not report what platform is used to recruit the evaluators. 
Among the remaining 18 papers, 11 use Amazon Mturk, 2 use Prolific, and 1 uses Microsoft UHRS, while 4 papers mention crowdsourcing platforms without specifying which one is used.

\textbf{Language background and geographic location of the evaluators:}
We find that 61.3\% of the papers we survey do not report whether the evaluators are native speakers of the language used in the speech synthesis model to be evaluated.
Furthermore, we observe that only 9 papers report the current location of their evaluators.
This presents a problem since the rating of native speakers and non-native speakers may differ, and the same language spoken by people from different parts of the world can also vary.


\textbf{Qualification of the evaluators:}
There is a possibility that even if the evaluator is a native speaker and resides in the region of interest, they may not be able to provide reliable feedback due to factors such as low-quality audio devices. 
It is also possible that the evaluator just wants to make money by answering the survey randomly.
Therefore, it is crucial to establish certain qualifications to filter out invalid evaluators and ensure the quality of the subjective evaluation. 
However, we note that a concerning number of papers (68 papers) do not address how they establish qualifications to select workers or handle invalid responses during post-processing.

\textbf{Instructions given to the evaluators:}
Task instructions serve to inform evaluators about the tasks at hand and provide guidance on how to complete the task.
In the MOS test, the instructions include the description used to describe a particular score, e.g., "\textit{5: Excellent}".
In our survey, two-thirds of the papers (51) fail to include any instructions used during their subjective evaluations. 
Many papers simply state that they "conduct a MOS test," without providing further details.
Although the recommended practice for MOS tests exists~\cite{rothauser1969ieee, rec1996p}, it is unclear whether the papers adhere to the evaluation procedures outlined in the recommendations.
In fact, we have observed the task instructions stated in some papers to be different from the recommendations.
We even find some papers (9) use a 0.5-point increment in the MOS tests, contradicting the 1-point increment in the recommended practice MOS tests.

\textbf{Number of raters and rated items:}
About one-third of the papers we survey do not report how many unique individuals participate in the subjective evaluation, and 27.5\% of papers do not say how many audio samples are evaluated.
More than half of the papers (51) do not state how many raters evaluate each audio sample, and 72 papers do not say the total number of audio samples rated by a unique individual.

\section{Experiment Setup}
\label{section: Experiment Setup}
We demonstrate the crucial role of unspecified details in subjective evaluation by conducting various MOS tests to evaluate the quality of three TTS models: Tacotron2~\cite{https://doi.org/10.48550/arxiv.1712.05884}, FastSpeech2~\cite{https://doi.org/10.48550/arxiv.2006.04558}, and VITS~\cite{https://doi.org/10.48550/arxiv.2106.06103}.
By manipulating certain factors in each MOS test, we investigate whether the experiment results vary. 
TTS is chosen as the target task since the majority of our surveyed papers focus on it, and we choose the three TTS models since they are well-studied and their performance is well-recognized.
\textbf{Since all the MOS tests share the same audio samples, there should only exist one ranking on the quality of the three TTS models, which is the ground truth ranking.}
Here, we do not assume what this ground truth ranking is, while there might be some agreement about this ranking in the TTS community.

\subsection{TTS Models and Datasets}
\label{subsection: TTS models}
We use LJSpeech~\cite{ljspeech17} as our dataset, which is commonly used in TTS research. 
For the TTS models, we use the pre-trained checkpoints from ESPNet-TTS\cite{hayashi2020espnet} and directly apply its demo code to synthesize all the samples. 
For FastSpeech2 and Tacotron2, we use the HifiGAN\cite{https://doi.org/10.48550/arxiv.2010.05646} vocoder checkpoint from ESPnet-TTS to convert the output spectrogram back to the waveform. 
All audios used in the experiment, including the ground truth audios, are normalized to mitigate the amplitude difference between speeches generated from different systems.

\subsection{Subjective Evaluation Setup}
\label{subsection: Subjective evaluation setup}
We randomly select 50 sentences from the testing set of LJSpeech and use the three TTS models to synthesize the corresponding audio samples.
The audio samples have lengths longer than 3 seconds and shorter than 10 seconds.
Each of the 50 sentences will have three audio samples generated by three TTS models and one human recording, resulting in a total of 200 audio samples.
We split the 200 audio samples into 10 equal-sized non-overlapping groups to form 10 questionnaires, and each questionnaire consists of 5 audio samples from the three TTS models and the human recordings.
There will be no audio samples in a questionnaire that have the same transcript.
Each audio sample is evaluated by 9 distinct evaluators.

Unless specified, we use the following instructions and rating scales in our MOS tests, following~\cite{maniati22_interspeech}.
We ask the evaluators "\textit{How natural (i.e. human-sounding) is this recording from a scale of 1 to 5?}".
The scale options are: \textit{"1: Bad - Very unnatural speech", "2: Poor - Somewhat unnatural speech", "3: Fair - Neither natural nor unnatural speech", "4: Good - Somewhat natural speech", "5: Excellent - Completely natural speech".}
We also ask the raters to wear headphones, and we only recruit workers that do not have hearing impairments.

We mainly use two crowdsource platforms for our experiments: Amazon Mturk and Prolific.
When using Amazon Mturk for evaluation, we cannot control the number of participants and how many audio samples an individual assesses.
We estimate that conducting a single questionnaire should take less than 5 minutes, and we pay the evaluators on Mturk US\$0.9 for conducting one questionnaire.
For the experiments conducted on Prolific, we recruit 9 distinct individuals and ask each of them to conduct the rating of 200 audio samples (10 questionnaires).
The interface seen by evaluators recruited from Prolific is the same as that seen by the workers recruited using Mturk.
Each individual is paid US\$10 for the rating of 200 audio samples, which is slightly higher than the payment to workers on Mturk. 
This is because workers on Prolific need to register a Mturk account to conduct the task, and we pay them slightly higher for doing so.
In all our subjective evaluations, we ensure that the payments are reasonable to the raters from anywhere in the world.
Other details about the experiments will be specified in the following sections.

In all the tables of our paper, we use subscripts to denote the width of the 95\% confidence interval of the MOS, and we use \colorbox{cyan!30}{blue}, \colorbox{yellow!30}{yellow}, and \colorbox{red!30}{red} to denote the best, runner-up, and worst TTS model.

\section{Do Different Factors in MOS Evaluation Affect the Result?}
In this section, we vary the factors in the MOS test and show that all these factors can change the experiment results.
\subsection{Qualification of Evaluators}
First, we study how the MOS test results can vary due to how we select the quality of the workers on Mturk.
In this section, we conduct our study on Mturk as it is the most adopted crowdsourcing platform in the papers we survey and it is a well-studied crowdsourcing platform~\cite{kittur2008crowdsourcing, ribeiro2011crowdmos}.
Mturk has two parameters to assess the quality of the workforce: \textbf{HIT Approval Rate} and \textbf{Number of HITs Approved}.
The former is the percentage of successfully completed tasks by a worker, while the latter represents the total number of completed tasks.
A higher HIT Approval Rate and Number of HITs Approved may indicate that the worker provides results with better quality.

We conduct two sets of MOS evaluation: the first one allows all the workers on Mturk to participate in the task and the second one only recruits workers that have HIT Approval Rate $\geq95\%$ and Number of HITs Approved $\geq 1000$; these numbers are set based on prior works that conduct human evaluations~\cite{karpinska-etal-2021-perils}.
For the MOS evaluation experiment in this section, we do not impose any additional requirements on the evaluators including geographic location and language background.

The results are presented Table~\ref{tab:qualification}. 
We show that without any worker qualifications (denoted as \textit{None} in Table~\ref{tab:qualification}), FastSpeech2 is favored over Tacotron2 in the MOS test. 
However, the highly overlapped 95\% confidence intervals of the MOS for the two models indicate that there is no statistical significance in FastSpeech2's superiority over Tacotron2. 
With a reasonably high worker threshold (i.e., HIT Approval Rate $\geq95\%$ and Number of HITs Approved $\geq1000$), the evaluators once again find Tacotron2 to be worse than FastSpeech2. 
Additionally, it seems that qualified listeners cannot distinguish between VITS and the ground truth. 
Based on these results, we will conclude that (1) although Tacotron2 is an autoregressive TTS model, the audio it synthesizes is still inferior to the audio samples produced by the non-autoregressive FastSpeech2, and (2) VITS is already on par with human recordings.

\begin{table}[th]
    \centering
    \caption{MOS results when using different qualifications. 
    }
    \label{tab:qualification}
    \begin{tabular}{c|ccc}
        \hline
        Qualification & None & $\geq$95\% and $\geq$1000 & Pass test \\
        \hline
        FastSpeech2 &\cellcolor{yellow!30} $3.70_{0.10}$ & \cellcolor{yellow!30}$3.70_{0.08}$ & \cellcolor{red!30}$3.17_{0.11}$\\
        Tacotron2  & \cellcolor{red!30} $3.62_{0.09}$ &  \cellcolor{red!30}$3.61_{0.08}$ & \cellcolor{yellow!30}$3.51_{0.10}$\\
        VITS  &\cellcolor{cyan!30} $3.78_{0.08}$ & \cellcolor{cyan!30} $3.74_{0.08}$ &  \cellcolor{cyan!30}$3.96_{0.10}$\\
        Ground truth  & $3.86_{0.08}$ & $3.74_{0.08}$ & $4.16_{0.08}$\\
        \hline
    \end{tabular}
    
\end{table}

Next, we ask whether we can use a test to filter valid evaluators and only recruit those workers passing the test to conduct the MOS test.
Using a test to select valid participants is recommended by P.808~\cite[Section 6.3.1.1]{rec2018p}, but it is unclear if this recommendation is widely adopted when conducting crowdsourcing subjective evaluations.
We design the test by the following procedure: 
We randomly sample 10 sentences in the test set of LJSpeech and synthesize 4, 3, and 3 audios using FastSpeech2, Tacotron2, and VITS, respectively.
Those sentences are different from the ones used for MOS tests.
We then pair those synthesized audios with the ground truth recording to form 10 pairs of audios.
Last, we create a survey containing the 10 audio pairs, and the participants are asked to choose the more natural sample in each pair of samples.
We publish the survey on Mturk and recruit 90 workers with HIT Approval Rate $\geq95\%$ and Number of HITs Approved $\geq1000$ to conduct the task, and they are paid for US\$0.9 for completing the survey.

\begin{figure}[ht!]
\centering
\includegraphics[clip, trim = 0px 40px 0px 52px,width=0.95\linewidth]{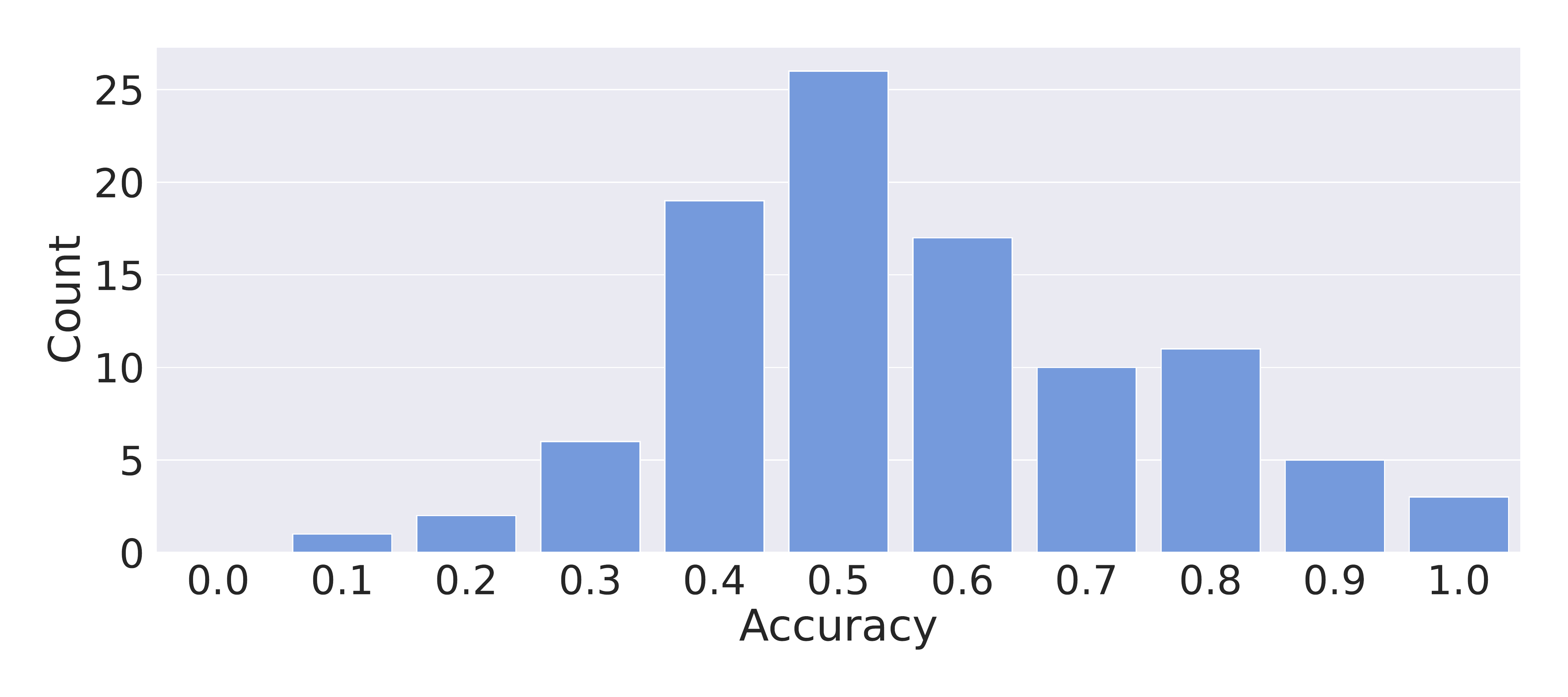}
\caption{The distribution of test accuracy.}
\label{fig:accuracy.pdf}
\end{figure}

We show the accuracy of the test in Figure~\ref{fig:accuracy.pdf}, where the accuracy is the proportion that a rater considers the ground truth to be more natural among the 10 audio pairs.
Surprisingly, more than half of the workers do not display a consistent preference for human recordings.
This finding suggests that setting qualifications on Mturk alone may not be sufficient if researchers expect evaluators to discern differences between model-generated and human recording samples.
We then conduct another MOS test while only allowing the workers with accuracy higher than 0.7 to participate, amounting to 29 workers.
The MOS test result, denoted with \textit{Pass test} in Table~\ref{tab:qualification}, reveals that VITS is the best, while Tacotron2 performs better than FastSpeech2.
The MOS differences between the three TTS models are all statistically significant.
This result contradicts our previous results.
Overall, qualifications  employed in the subjective evaluation may result in a selection bias on the experiment result. Therefore, it is crucial to report the qualifications used.

\subsection{Location of Workers}
Next, we study how the locations of workers change the MOS results using Mturk.
We only recruit English speakers as they are more familiar with English and hence may be better equipped to detect subtle unnatural prosody or accent in the samples. 
However, Mturk assumes that workers using their platform are fluent in English; therefore, no qualification for the English ability of the raters can be set.
We publish three MOS tests on Mturk, recruiting only workers from the USA, the UK, and India respectively.
We also only recruit workers that have HIT Approval Rate $\geq95\%$ and Number of HITs Approved $\geq1000$.

The experiment results are shown in Table~\ref{tab:location}.
We find that for workers in the USA, FastSpeech2 generates audio samples as natural as those generated by Tacotron2.
Workers in India also agree that the quality of FastSpeech2 and Tacotron2 is very similar.
However, raters in the UK consider Tacotron2 superior to FastSpeech2 by a significant margin.
Furthermore, UK-based evaluators consider VITS much more unnatural compared to the ground truth, while workers in the other two regions do not find the ground truth significantly better. 
We include the result when we do not restrict the location of the raters in Table~\ref{tab:location}, denoted as \textit{All}.
In this case, we observe a completely different ranking among the three TTS models.
This highlights the variability of the results due to the location of the evaluators.

\begin{table}[ht]
    \centering
    \caption{MOS results when recruiting evaluators residing in different locations. 
    }
    \label{tab:location}
    \begin{tabular}{c|cccc}
        \hline
        Location & All & USA & UK & India \\
        \hline
        FastSpeech2 & \cellcolor{yellow!30}$3.70_{0.08}$ &\cellcolor{yellow!30} $3.73_{0.09}$&\cellcolor{red!30} $2.64_{0.08}$&\cellcolor{red!30} $3.58_{0.09}$\\
        Tacotron2  &\cellcolor{red!30} $3.61_{0.08}$ & \cellcolor{yellow!30}$3.73_{0.09}$ & \cellcolor{yellow!30}$2.87_{0.09}$&\cellcolor{yellow!30} $3.62_{0.08}$\\
        VITS  & \cellcolor{cyan!30} $3.74_{0.08}$& \cellcolor{cyan!30} $3.79_{0.09}$& \cellcolor{cyan!30}$3.17_{0.09}$& \cellcolor{cyan!30}$4.10_{0.07}$\\
        Ground truth & $3.74_{0.08}$  & $3.87_{0.08}$& $3.71_{0.08}$& $4.15_{0.07}$\\
        \hline
    \end{tabular}
    
\end{table}

The phenomenon observed in this section could be attributed to several potential factors.
From a linguistic perspective, English spoken by speakers from different regions could vary, potentially affecting how raters score the same audio sample.
Another possible reason could be that people from the USA are more tolerant of unnatural samples, resulting in them rating samples as more natural.
Additionally, the headphones used by evaluators from different countries may be systematically different, leading to different perceptions of the unnatural elements in the audio samples.
There could be more intricate reasons that are not listed here, and all of them contribute to the uncertainty of subjective evaluation results. 
Thus, it is important to report the locations of evaluators who participated in the study to better understand to whom the experiment results may apply.

\subsection{Crowdsourcing Platforms}
In this section, we turn our attention to the crowdsourcing platform used to recruit evaluators.
We choose two popular platforms, Mturk and Prolific, and recruit workers located in the USA for both platforms. 
We also publish another MOS test by recruiting students enrolled in a Machine Learning course at our university to conduct the study.
The demographic constitution of the raters recruited at our university is significantly different from the workers on Mturk and Prolific: students participating in our study are Asian whose first language is Chinese but can speak English fluently; the age distribution of the students falls in the range of 18 to 28.
We include the study using students from our university because it is common for graduate student researchers to conduct subjective evaluations using their personal networks, and we aim to simulate this scenario by recruiting students on campus.

The results are presented in Table~\ref{tab:platform}.
Even though the demographic composition of the workers recruited from Prolific is markedly different from that of our university, they produce the same ranking of the TTS models. 
However, evaluators on Prolific are more adept at distinguishing the quality disparity between samples generated by FastSpeech2 and Tacotron2. 
In contrast, workers from Mturk do not find significant differences in the quality of samples produced by the three TTS models.

\begin{table}[ht]
    \centering
    \caption{MOS results when recruiting evaluators using different platforms. 
    }
    \label{tab:platform}
    \begin{tabular}{c|ccc}
        \hline
        Platform & Mturk & Prolific & University \\
        \hline
        FastSpeech2 & \cellcolor{yellow!30} $3.73_{0.09}$ & \cellcolor{red!30} $2.81_{0.11}$ & \cellcolor{red!30} $3.08_{0.12}$\\
        Tacotron2  & \cellcolor{yellow!30} $3.73_{0.09}$ & \cellcolor{yellow!30} $3.02_{0.11}$ & \cellcolor{yellow!30} $3.18_{0.12}$\\
        VITS  & \cellcolor{cyan!30} $3.79_{0.09}$ & \cellcolor{cyan!30} $3.12_{0.11}$ & \cellcolor{cyan!30} $3.46_{0.11}$\\
        Ground truth  & $3.87_{0.08}$ & $4.12_{0.08}$ & $3.76_{0.11}$\\
        \hline
    \end{tabular}
    
\end{table}

The possible reasons for the result differences are discussed as follows:
Different recruiting platforms have different processes for how to become a valid worker on the platform.
For instance, Prolific necessitates that workers verify their phone numbers and government ID, while Amazon Mturk may not mandate the provision of government ID by workers.
These differences may potentially affect the quality of the workforce by serving as a prescreening mechanism.
Secondly, the number of unique raters involved in the studies conducted on different platforms is different, which may potentially affect the results. 
In this section, the studies conducted on Mturk, Prolific, and our university involved 90, 9, and 90 unique participants, respectively.
The impact of the unique number of raters on the experiment results will be investigated  with more systematic analyses in future work.
Although we only controlled the crowdsourcing platform in this section, numerous factors can change by simply altering the crowdsourcing platform. 
Since the platform can significantly influence the experiment results, it is crucial to explicitly state the platform used to help readers better understand the potential underlying distribution of evaluators in the study.

\subsection{Instructions to the Workers}
Last, we investigate how the MOS results can change by varying the instructions given to the workers.
The experiments in this section are conducted on Prolific and only recruit workers living in the USA whose first language is English.
We use four sets of instructions to create four different MOS experiments, and the workers in all four experiments are non-overlapping.
The instructions are: 
(i) \textbf{None}: \textit{"How natural (i.e. human-sounding) is this recording on a scale of 1 to 5? 1: Poor, 2: Bad, 3: Fair, 4: Good, 5: Excellent."} 
This follows the P.800~\cite[B.4.5]{rec1996p}.
(ii) \textbf{Natural}: The default instruction stated in Section~\ref{subsection: Subjective evaluation setup}.
(iii) \textbf{Distort}: \textit{"What is the quality of the speech based on the level of distortion of the speech on a scale of 1 to 5?  1: Bad - Very annoying and objectionable, 2: Poor - Annoying, but not objectionable, 3: Fair - Perceptible and slightly annoying, 4: Good - Just perceptible, but not annoying, 5: Excellent - Imperceptible."} This follows the MOS (ACR) referred to in~\cite{ribeiro2011crowdmos}.
(iv) \textbf{All}: We use the default instruction in Section~\ref{subsection: Subjective evaluation setup}, but explicitly instruct the raters to consider the \textit{"fluency, prosody, intonation, distortion, and noise in the sample."}
This instruction is motivated by 2 papers in our survey that explicitly instruct the evaluators on what to focus on during the evaluation.

\begin{table}[t]
    \centering
    \caption{MOS results when using different task instructions.
    We also report the average time an evaluator takes to complete the rating of 200 samples.
    }
    \label{tab:intrstruction}
    \begin{tabular}{c|cccc}
        \hline
        Instruction & None & Natural & Distort & All \\
        \hline
        FastSpeech2 &\cellcolor{red!30} $3.11_{0.1}$ & \cellcolor{red!30}$3.06_{0.1}$ &\cellcolor{red!30} $3.0_{0.09}$ & \cellcolor{yellow!30} $2.96_{0.1}$\\
        Tacotron2  &\cellcolor{yellow!30}$3.16_{0.1}$ & \cellcolor{cyan!30} $3.23_{0.1}$ & \cellcolor{yellow!30}$3.20_{0.1}$ &  \cellcolor{cyan!30}$3.10_{0.1}$\\
        VITS  &  \cellcolor{cyan!30} $3.40_{0.12}$ & \cellcolor{yellow!30} $3.14_{0.11}$ &  \cellcolor{cyan!30}$3.98_{0.1}$ & \cellcolor{red!30}$2.95_{0.11}$\\
        Ground truth  & $4.28_{0.08}$ & $3.96_{0.09}$ & $4.57_{0.07}$ & $3.89_{0.08}$\\
        \hline
        Time (mins) & 32 & 43 & 52 & 52\\
        \hline
    \end{tabular}
    
\end{table}

The results in Table~\ref{tab:intrstruction} show three different rankings of the three TTS models.
With the \textbf{None} instruction containing the least instruction, raters find VITS to be the best TTS model, with the shortest time taken to complete the task among the four settings.
When using the default instruction (\textbf{Natural}), Tacotron2 becomes the best one.
When raters are asked to focus on the distortion in the samples (the \textbf{Distort} instruction), the raters again agree that VITS has the least distortion. 
We find that VITS becomes the worst TTS model for the raters when they are asked to consider all possible factors for natural speech using the \textbf{All} instruction.
We also observe that when the instructions are longer, the time taken to complete the task becomes longer. 
Additionally, when the evaluators are explicitly asked to focus on certain factors in the samples (as in \textbf{Distort} and \textbf{All}), they spend more time on the task. 
After finishing the task, we interview the participants in the \textbf{None} group and ask them what factors they consider during the rating. 
Interestingly, they state that fluency, pronunciation, robotic sounds (distortion), and noises are the main factors, which mostly coincide with the factors we listed in the \textbf{All} setting. 
This shows that even when the raters consider similar factors during the tasks, the results can still be largely different depending on whether they are explicitly required to do so.

\section{Conclusion}
In this paper, we reveal that most papers on speech synthesis do not fully report the details of subjective evaluations.
To highlight the gravity of the problem, we conduct more than ten sets of MOS experiments to rate the quality of three TTS models and obtain at least three rankings on the quality of those models.
Since all the MOS evaluation shares the same audio samples but only differ in the factors in subjective evaluation, we show that those factors are highly influential to the experiment results.
The surveyed paper list and the example of MOS tests can be found at github.com/d223302/SubjectiveEvaluation.
Since we do not assume a ground truth ranking of the TTS models used in our paper, we are not able to provide any guidelines on how to conduct "\textit{better}" subjective evaluations to yield results closer to the ground truth. 
The one and only guideline we provide for future researchers when conducting \textit{good} subjective evaluations is to comprehensively report every detail in the subjective evaluations.
While there are guidelines for conducting crowdsourcing MOS evaluation~\cite{kittur2008crowdsourcing,ribeiro2011crowdmos}, it is unclear if those guidelines are still adopted recently and if they are suitable nowadays.

While the details in human evaluation have been included in the checklist of major machine learning and natural language processing conferences (e.g., NeurIPS and *ACL), the speech community has yet to take similar action. 
To increase the reproducibility of experiment results and allow for more reliable interpretations of subjective evaluation results, we encourage future researchers to comprehensively report details in subjective evaluations, either in the paper or by online supplementary materials.
We hope that the concerning results presented in our paper draw attention to the importance of reporting subjective evaluation details and provoke further discussions on this topic.

\bibliographystyle{IEEEtran}
\bibliography{mybib}

\end{document}